\documentclass[9pt,twocolumn,twoside]{opticajnl}
\journal{opticajournal} 

\setboolean{shortarticle}{false}

\usepackage{ifthen}
\usepackage{wrapfig}
\usepackage{lineno}
\usepackage{upgreek}
\usepackage{xfrac}
\usepackage{array}
\usepackage{gensymb}

\title{C2PO: Coherent Co-packaged Optics using offset-QAM-16 for Beyond PAM-4 Optical I/O}

\author[1,*]{Dan Sturm}
\author[1]{Marziyeh Rezaei}
\author[1]{Alana Dee}
\author[1]{Sajjad Moazeni}

\affil[1]{Department of Electrical and Computer Engineering, University of Washington, Seattle, WA, 98195, USA}

\affil[*]{dansturm@uw.edu}
\begin{abstract} 
Co-packaged optics (CPO) has emerged as an ultimate solution for achieving the ultra-high bandwidths, shoreline densities, and energy efficiencies required by future GPUs and network switches for AI. Microring modulators (MRMs) are well-suited for transmitters due to their compact size, high energy efficiency, and natural compatibility for dense wavelength-division multiplexing (DWDM). However, extending beyond the recently-demonstrated 200 Gb/s will require more advanced modulation formats such as higher-order coherent  (e.g., QAM-16).
\\ 
In this work, we show how microring resonators (MRMs) can be efficiently used to implement phase-constant amplitude modulators and form the building blocks of a transmitter for offset-QAM-16, which has been shown to simplify carrier-phase recovery relative to QAM with no offset. We simulate and evaluate the performance of our proposed MRM-based coherent CPO (C2PO) transmitters using a foundry-provided commercial silicon photonics process, demonstrating an input-normalized electric field amplitude contrast of 0.64 per dimension. Through full link-level bit error rate modeling, we show that our design achieves 400 Gb/s using offset-QAM-16 at a total optical laser power of 9.65 dBm—comparable to that required by conventional QAM-16 MZI-based links—despite using 10–100× less area. We further conduct a thermal simulation to assess the transmitter’s thermal stability at the MRM input optical power required to meet a target BER at the desired data rates. Finally, as a proof of concept, we demonstrate 25 Gb/s MRM-based offset-QAM-4 modulation with a chip fabricated in the GlobalFoundries 45 nm monolithic silicon photonics process.
\newline
\end{abstract}

\setboolean{displaycopyright}{false} 

\begin{document}

\maketitle

\section{Introduction and Related Work}

\begin{figure}
    \centering
    \includegraphics[width=0.85\linewidth]{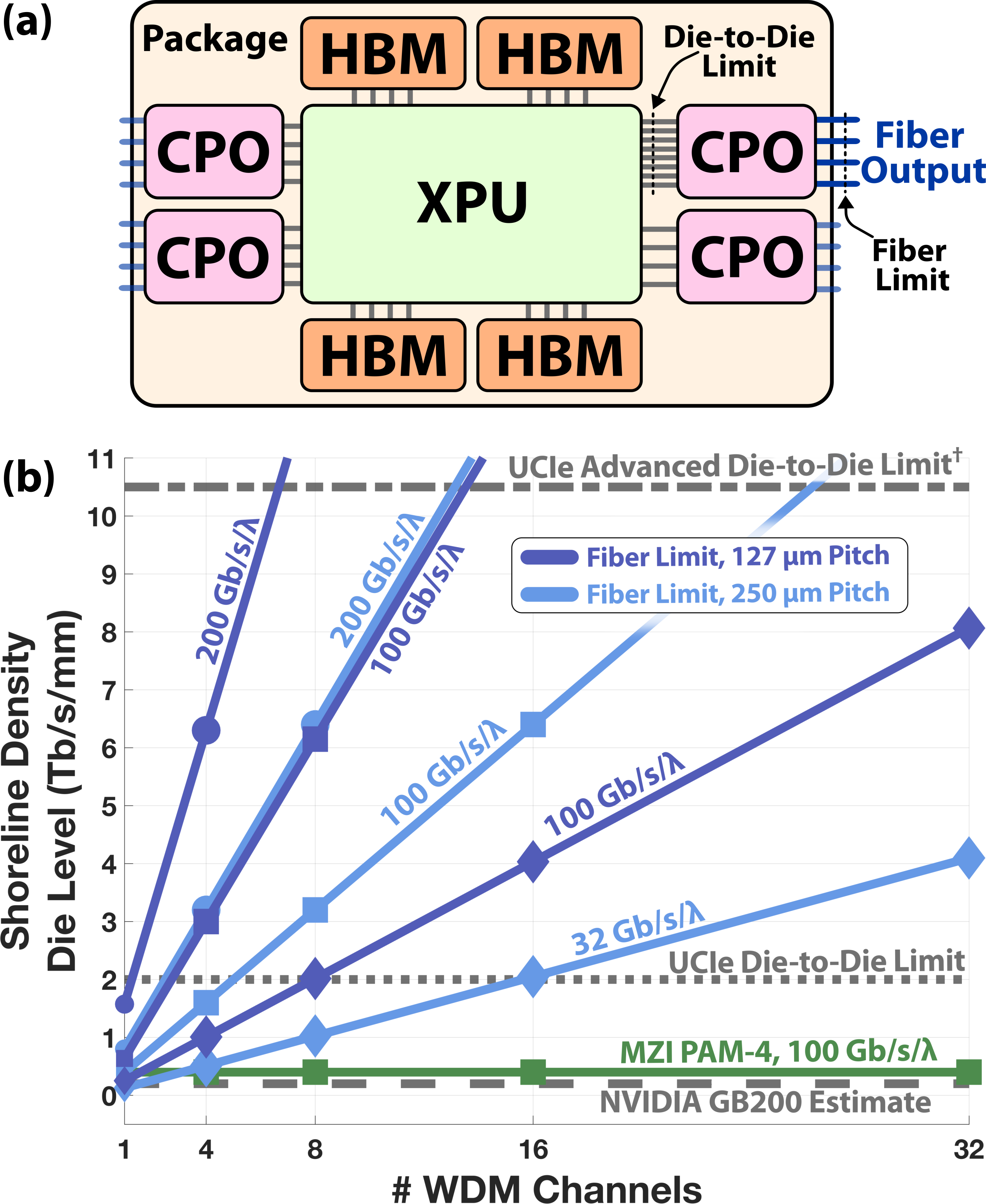}
    \caption{(a) Illustration of future System-in-Package (SiP) with CPO for optical I/O, and (b) die-level shoreline density comparison and requirements for CPO solutions at standard fiber pitches of 127 or 250 $\upmu$m.\newline$\dag$The UCIe 1.0 Advanced spec is for 2.5D packaging, i.e. dies connected with a silicon bridge or interposer~\cite{ucie}}
    \label{fig:cpo_shoreline}
\end{figure}

As AI/ML accelerators, GPUs, and network switches increasingly demand tens of Tb/s of off-package IO bandwidth at pJ/bit efficiencies and 1+ Tb/s/mm shoreline densities, the adoption of co-packaged optics (CPO) is becoming inevitable~\cite{intel_CPO_full}. Although today's pluggable optical modules use Mach-Zehnder Modulator (MZM)-based transmitters~\cite{he202156}, their large footprint (multi-mm) bottlenecks shoreline density for CPO and prevents them from reaching the limits imposed by fiber pitch. Hence, CPO will rely on microring modulators (MRMs) as electro-optical transmitters. While the compact footprint of these devices (10s of $\upmu$ms) enables high density, overall shoreline density will still be limited by the pitch of fibers used to take optical signals off package (Fig~\ref{fig:cpo_shoreline}a, b). Thus, the future of CPO and shoreline density scaling will rely on increasing the datarate per fiber of MRM-based transmitters.

MRM-based transmitters currently utilize only amplitude modulation, such as pulse amplitude modulation (PAM) (e.g., NRZ and PAM-4). Wavelength division multiplexing (WDM) has been proposed to achieve higher shoreline densities and is naturally suitable for MRMs due to their wavelength selectivity~\cite{Winzer:23}. However, with conventional MRM datarates of 32 or even 100 Gb/s/lane, system performance will be limited by the large number of WDM lines required for meeting target densities, given the challenges of making high-quality multi-wavelength laser sources and managing the thermal sensitivity of MRMs (Fig.~\ref{fig:cpo_shoreline}b). Meanwhile, scaling with polarization division multiplexing (PDM) is also plausible, however this is limited by physics and also requires more expensive polarization maintaining (PM) fibers and assembly. Therefore, reaching 200 or 400 Gb/s per wavelength is becoming a must to meet the future shoreline densities provided by advanced die-to-die interconnects like 3D UCIe~\cite{UCIe-NatElec2024}. However, scaling the baud rate beyond 100 GBd is extremely challenging as CMOS scaling no longer improves the mixed-signal circuit performance. Achieving energy-efficient higher data rates per lane to increase shoreline density will thus require more advanced coherent modulations.

Building MRM-based QAM-16 transmitters can increase the data rate to 200 or 400 Gb/s without using more laser lines or increasing the baud rate, and can thus be an ultimate solution for future CPO. Although coherent optics using quadrature amplitude modulation (QAM) has been widely used for long-haul (+10km) applications with Mach-Zehnder modulator (MZM)-based transmitters~\cite{valenzuela202250}, it is extremely challenging with MRMs due to the coupled nature of their phase and amplitude outputs. 

Coherent MRM-based transmitters have previously been demonstrated in the form of single-ring BPSK modulators, which modulate the MRM on either side of its resonance to induce a $\pi$ phase shift~\cite{mehta_laser-forwarded_2019, jo_novel_2024, debnath_generation_2024, dong_silicon_2013, padmaraju_error_free_2011, padmaraju_dpsk_2011}. 
These approaches are limited to two output levels per dimension and thus cannot be used for higher-order QAM modulation. Another set of approaches uses multiple MRMs and/or couplers to modulate amplitude and phase separately but these introduce high modulation loss making them unsuitable for high-bandwidth, low-BER links \cite{one_ring_amp_one_ring_phase, multi_coupling_1, multi_coupling_2, multi_coupling_3}.

To address these shortcomings, we propose a novel offset-QAM-16 transmitter architecture using Ring-Assisted MZI (RAMZI)~\cite{OptExp2013-RAMZI} structures. Previous works have introduced the RAMZI configuration to achieve low-chirp PAM modulators~\cite{ramzi_David_plank, ramzi_google}, but do not demonstrate the RAMZI's application for coherent communication (i.e. QAM-16). RAMZIs have been demonstrated for coherent communications in \cite{laval_main, laval_small}, but these approaches rely on individual MRMs being biased at the laser wavelength, which has been shown to significantly reduce bandwidth~\cite{MRM_tradeoff_1, MRM_tradeoff_2}.

In this work, we use frequency-detuned RAMZIs to generate phase-constant amplitude modulators, which are the building blocks of an offset-QAM transmitter capable of performing offset QAM-16 or higher-order coherent modulation. While offset has been typically suppressed for conventional QAM signaling, we show that the impacts of the offset will be minimal to OMA and bit errors induced by phase noise. In addition, we have recently exploited the offset signal to realize a DSP-free carrier phase recovery approach for QAM-16~\cite{marizyeh_recovery}. Our work uses the O-band (1310 nm) because its low dispersion makes it the most practical choice for future intra-datacenter optical links \cite{marvell_website_oband_coherent}. 

This work includes a full circuit-level electro-optical modeling of the proposed transmitter, which we verify via simulation in an advanced silicon photonics process (Sections~\ref{Proposed Circuit and Operation}, \ref{sec:simulation methodology}). We then perform a link-level bit error rate (BER) modeling to demonstrate that RAMZI-based links can acheive target datarates with similar total optical powers as MZM-based links (Section~\ref{sec:link modeling}). We also simulate and study the thermal stability of the MRMs for tuning and operation at the required optical power (Section~\ref{thermal analysis}). 

We have validated this method using the GlobalFoundries (GF) 45 nm monolithic silicon photonics process (GF45SPCLO)~\cite{ GF45SPCLO-OFC2020} and show results in Section~\ref{chip results}. In Section~\ref{comparison SoA} we compare the RAMZI-based offset-QAM-16 (ROQ) transmitter (ROQ-T) with state-of-the-art solutions and show that it can achieve similar bit efficiencies to conventional MZM-based links while consuming $\sim\!\text{10-100}\times$ lower area. We believe that this will enable the next generation of ultrafast, power- and area-efficient coherent CPO for realizing the vast potential of AI accelerators, GPUs, and network switches.

\section{Proposed MRM-based Coherent Transmitter} \label{Proposed Circuit and Operation}

\begin{figure} [b]
    \centering
    \includegraphics[width=0.9\linewidth]{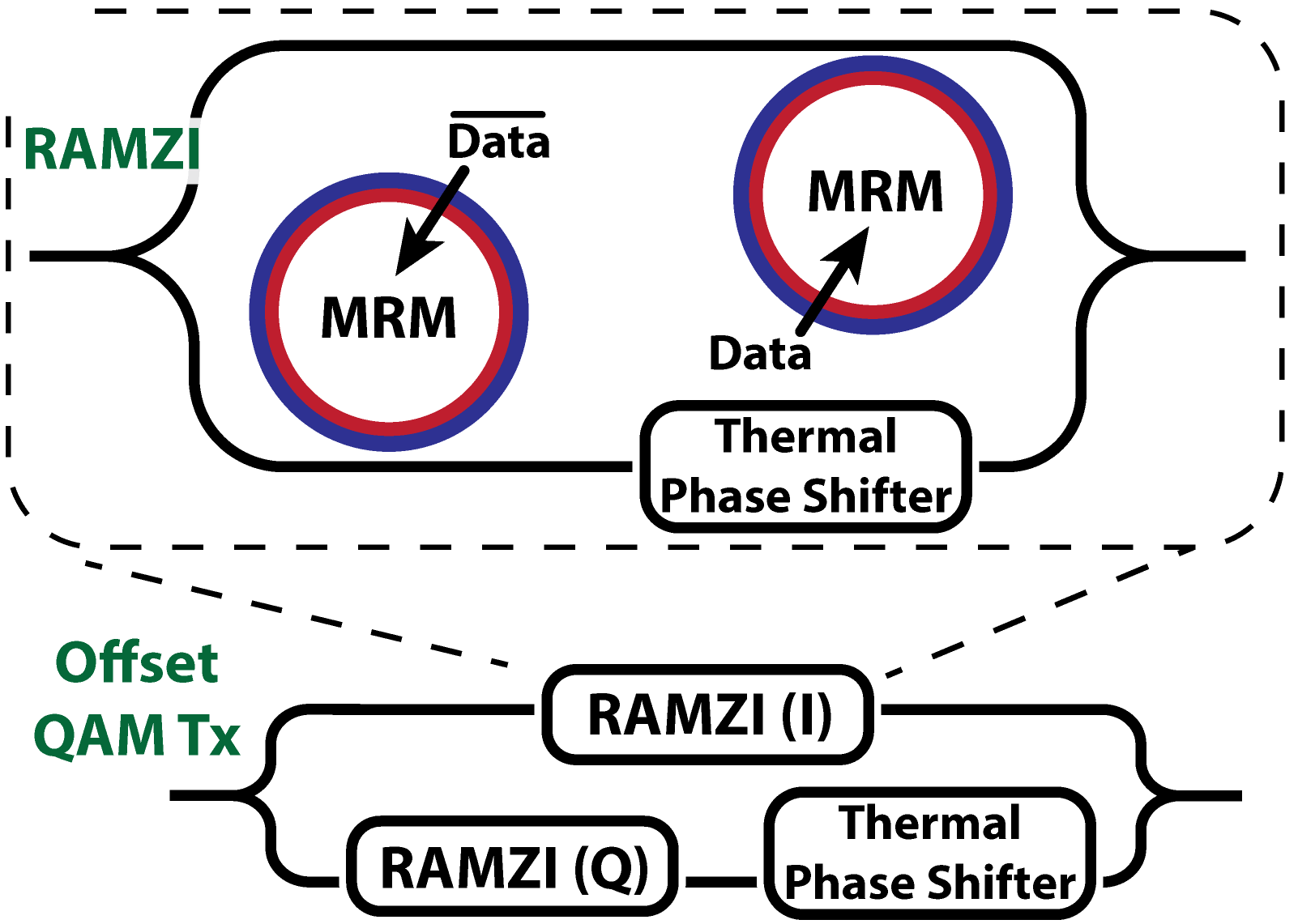}
     \caption{Proposed RAMZI-based offset-QAM Transmitter and RAMZI photonic circuit.}
    \label{fig:ramzi and qam tx}
\end{figure}

\begin{figure}[h]
    \centering
    \includegraphics[width=0.95\linewidth]{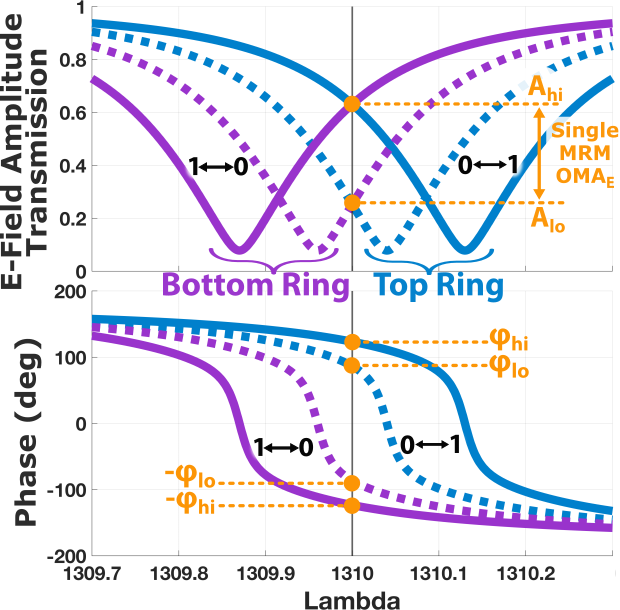}
    \caption{Principle of operation for phase-constant RAMZI modulator.}
    \label{fig:ramzi biasing}
\end{figure}

\begin{figure}[b]
    \centering
    \includegraphics[width=0.9\linewidth]{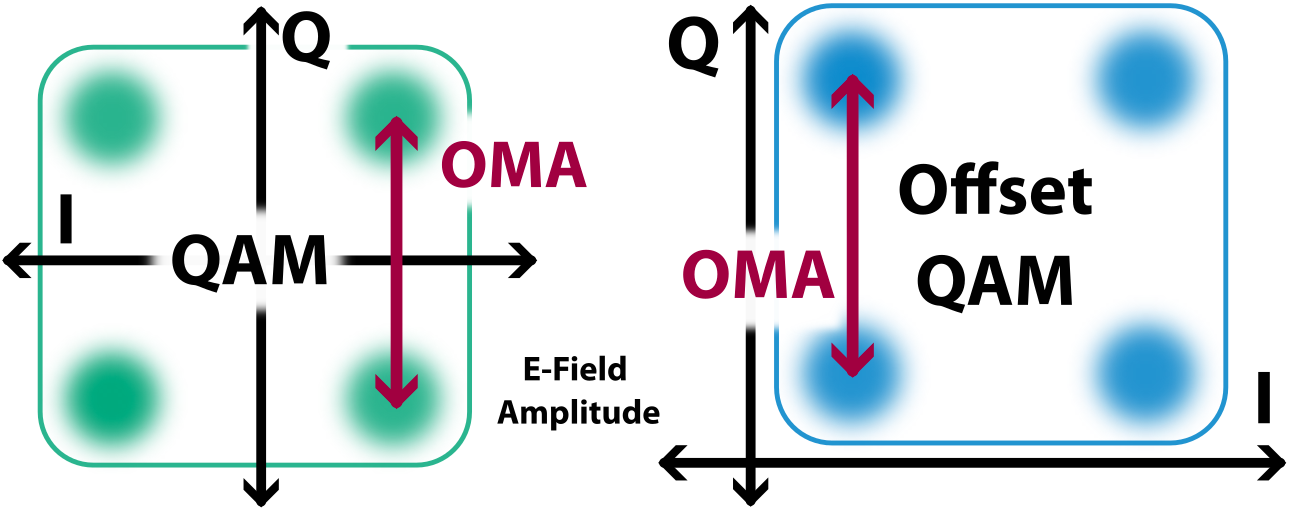}
    \caption{Offset-QAM and QAM Constellations. In our proposed RAMZI-based transmitter, the offset is tunable and can be zero, to enable regular QAM operation}
    \vspace{0.25cm}
    \label{fig:offset qam constellation}
\end{figure}

The proposed RAMZI photonic circuit consists of two MRMs on either arm of an MZI (Fig.~\ref{fig:ramzi and qam tx}). Each MRM's resonance wavelength can be thermally tuned to ensure that their resonances lie on opposite sides of the laser wavelength, and differential (i.e., push-pull) driving voltages are applied to simultaneously modulate the MRMs' resonances in opposite directions towards or away from the laser wavelength (Fig.~\ref{fig:ramzi biasing}). This gives each MRM an identical amplitude change with opposite phase shift, which modulates the RAMZI's output amplitude while holding output phase constant. A bias thermal phase shifter is included in one arm of the RAMZI. This can be tuned so that the individual MRM outputs interfere constructively in the high transmission state and destructively in the low transmission state. Thus, in addition to offering phase-constant operation, the RAMZI offers an amplitude contrast superior to that of a single MRM by taking advantage of both amplitude and phase modulation in the MRM (Eq.~\ref{Eq: ramzi oma}). Two RAMZIs with separate data inputs can be combined with a $\sfrac{\pi}{2}$ phase shift to make an offset-QAM transmitter (Fig.~\ref{fig:ramzi and qam tx}). 

Since the RAMZI has constant phase but variable power output, the constellation of a RAMZI-based transmitter is offset from the origin of the I-Q plane (Fig~\ref{fig:offset qam constellation}). Although unintuitive, the offset has been shown to be beneficial in realizing a DSP-free carrier phase recovery for QAM-16~\cite{marizyeh_recovery}, and has minimal impacts to impacts to OMA and the impact of phase error on bit error rate (Sec.~\ref{sec:simulation methodology}, \ref{sec:link modeling}). This offset is adjustable via tuning the bias points of each MRM to be closer to (less offset) or further from (more offset) the laser wavelength. When both MRMs are biased at the laser wavelength, QAM with no offset is performed. The offsets of I and Q branches can be tuned separately.

We let $A_Te^{i\phi_T}$ and $A_Be^{i\phi_B}$ be the output signal from the top and bottom MRMs, respectively, and let $\phi_{PS,\,Arm}$ be the phase shift of the thermal phase shifter after accounting for the $\sfrac{\pi}{2}$ phase delay of the bottom branch inherent to the output combiner. Since the MRMs always have opposite phases and equal amplitudes, we can represent $\phi_T$ and $\phi_B$ as $\phi_X$ and $-\phi_X$, and $A_T$ and $A_B$ as $A_X$. We calculate the output electrical field as: 
\begin{equation}
    E_{out} = \frac{1}{\sqrt{2}}(A_Te^{i\phi_T}+A_Be^{i(\phi_B-\phi_{PS})}) 
\end{equation}

We calculate RAMZI output phase ($\phi_{out,R}$) and power ($P_{out,R}$) as:
\begin{align}
\phi_{out,R} &= \tan^{-1}\left( 
    \frac{A_T \sin(\phi_T) + A_B \sin(\phi_B - \phi_{PS})}{A_T \cos(\phi_T) + A_B \cos(\phi_B - \phi_{PS})}
\right) \nonumber \\
&= \tan^{-1}\left( 
    \frac{A_X}{A_X} \times 
    \frac{\sin(\phi_X) + \sin(-\phi_X - \phi_{PS})}{\cos(\phi_X) + \cos(-\phi_X - \phi_{PS})}
\right) \nonumber \\
&= -\phi_{PS} \label{Eq:ramzi output phase} \\
P_{out,R} &\propto \frac{1}{2}(A_T^2 + A_B^2+ 2{A_T}{A_B}\cos(\phi_T - \phi_B + \phi_{PS})) \label{Eq: ramzi output power} \nonumber \\
&\propto A_X^2(1 + \cos(2\phi_X+\phi_{PS}))
\end{align}

\newcommand{\omae}{$\text{OMA}_\text{E}$}

Equation \ref{Eq:ramzi output phase} shows that output phase is constant. Since coherent optical receivers detect power proportional to the transmitted electric field strength, we calculate the effective optical modulation amplitude (\omae{}) of a device/circuit as the contrast between the output electric field amplitude in its high and low states. In this paper we will focus on the normalized \omae{}, in which all values are referenced to the RAMZI's electric field amplitude input. We let $A_{Hi}$, $A_{Lo}$, $\phi_{Hi}$, $\phi_{Lo}$ be the MRM electric field amplitude ($A_X$) and phase output ($\phi_X$) from Eqs.~\ref{Eq:ramzi output phase} and~\ref{Eq: ramzi output power}. We then calculate the RAMZI's \omae{} as: 
\begin{align}
    OMA_{E} &= \left| E_{out, Hi} \right| - \left| E_{out, Lo} \right| \nonumber \\
    &= A_{Hi}\sqrt{1+\cos(2\phi_{Hi}+\phi_{PS})}  \label{Eq: ramzi oma}\\
    &\quad\quad\quad - A_{Lo}\sqrt{1+\cos(2\phi_{Lo}+\phi_{PS})} \nonumber
\end{align}
The $\text{OMA}_\text{E}$ can be greatest when there is a large contrast between $\phi_{hi}$ and $\phi_{lo}$. Maximum MRM output phase contrast might not occur at the same bias point as maximum MRM amplitude contrast. Thus, the bias point that maximizes the RAMZI's \omae{} may not be the one that maximizes the top and bottom MRMs' \omae{}.

The inherent wavelength selectivity of MRMs makes the RAMZI-based offset-QAM transmitter (ROQ-T) naturally adaptable for WDM operation. A WDM RAMZI can be constructed by placing multiple pairs of MRMs (each with their own resonance wavelength) on either arm of the same MZI so that a single thermal phase shifter in one arm is shared between them. This amortizes the area of the phase shifter over multiple wavelengths (4, 8, etc) to minimize overall transmitter area and make the area of a single RAMZI comparable to two single MRMs (i.e. those used in PAM operation). 

While the coherent operation provided by ROQ provides a great benefit for intra-datacenter interconnect compared to MRM-based amplitude modulation, the underlying electro-optical devices (MRMs) the same. As a result, the existing peripheral electronics developed for MRM-based amplitude modulation can be used, facilitating ROQ adoption. This includes closed-loop thermal tuning of MRM resonance points with a drop port photodiode \cite{intel_thermal_tuning} as well as high-swing RF drivers for data modulation \cite{MRM_driver_intel}. We address thermally tuning MRMs on both sides of resonance in Sec.~\ref{thermal analysis}. 

\section{Simulation Methodology and Results}\label{sec:simulation methodology}
\begin{figure}[h]
    \centering
    \includegraphics[width=0.65\linewidth]{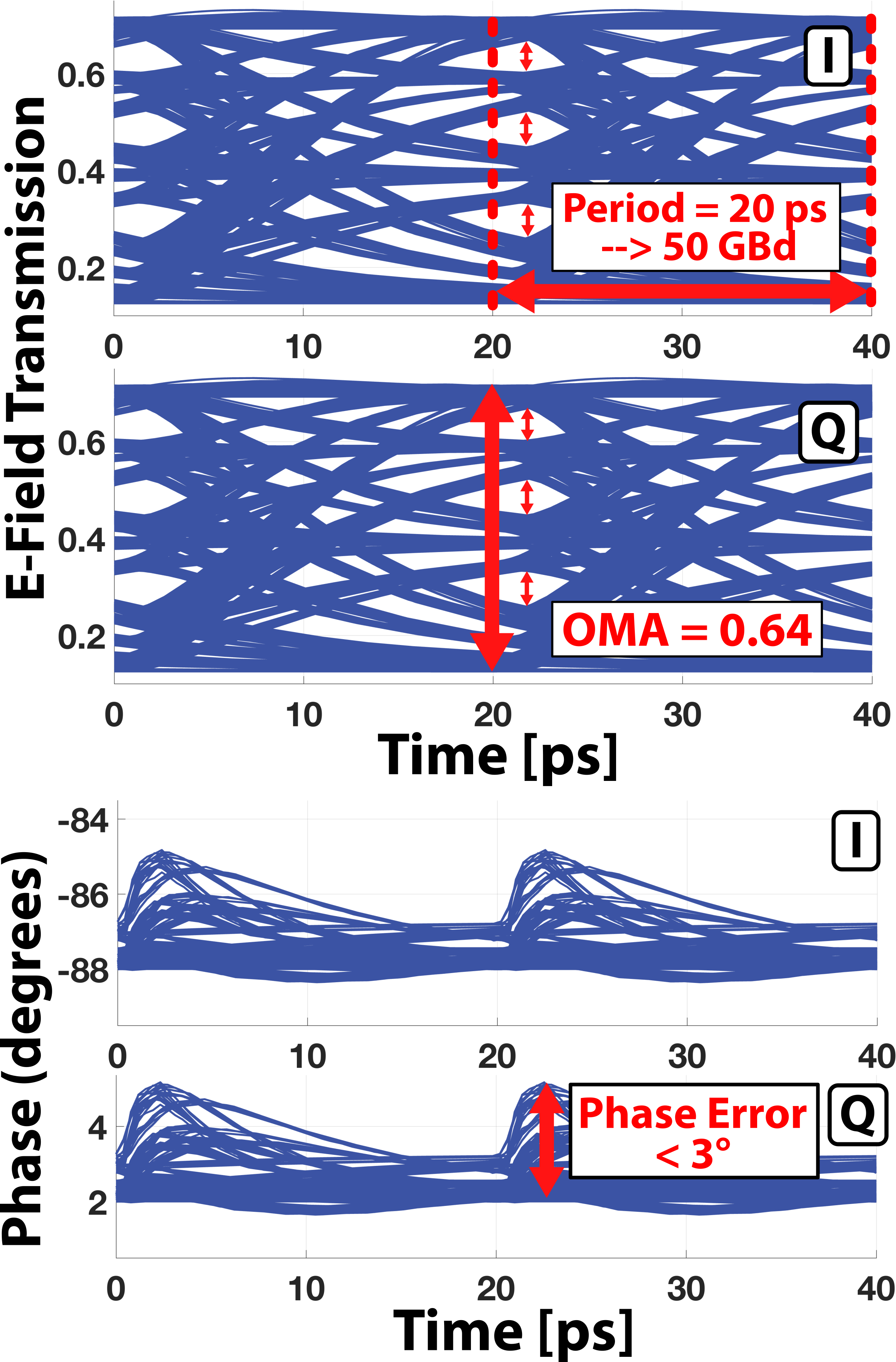}
    \caption{Simulated electric field (E-field) amplitude (a) and phase (b) eye diagrams for the proposed RAMZI-based offset-QAM transmitter at 50 GBd.}
    \label{fig:eye diagram}
\end{figure}

We simulate the performance of the proposed photonic circuit using Spectre PDK models of the GF45SPCLO process in Cadence Virtuoso. These models include an MRM with a radius of 7.5 $\mu$m and $\sim 35~GHz$ electro-optical bandwidth, as well as other photonic devices (e.g. thermal phase shifter, 1x2 MMI splitter, etc.). We use ideal 4-level PRBS voltage sources with 20\% rise and fall times to drive all MRMs. 

The first step is to find the correct thermal tuning levels for each MRM to maximize the RAMZI's \omae{} with each MRM resonance having an equal offset on opposite sides of the laser wavelength. We sweep the applied heat while applying a laser input in the O-band ($\sim$1310 nm) and 0V and -4V inputs to each MRM PN junction, which is the typical overall driving voltage swing for MRMs. By applying differential driving voltages to the MRMs, the RAMZI output amplitude is modulated while phase is held constant. For each combination of thermal tuning levels that give each MRM equal offsets on opposite sides of the laser wavelength, we sweep the RAMZI's arm phase shifter. Ultimately we pick the combination of MRM thermal tuning and RAMZI arm phase shift tuning that maximizes the RAMZI's \omae{}. Alternatively, this process could be done to optimize the RAMZI's offset level instead of the \omae{}, or a combination of the two. 

We then find the pairs of intermediate driving voltages for top and bottom MRMs ($V_{MRM,T}, V_{MRM,B}$) that yield phase-constant RAMZI output electric field strengths evenly spaced between the high and low outputs found previously. Due to the nonlinearity of the MRM response with respect to applied voltage, as well as the dependence of RAMZI output on both amplitude and phase modulation in each MRM, this must be done iteratively. We sweep through all values of $V_{MRM,T}, V_{MRM,B}$ between 0 and -4 to find the combinations that yield equal amplitude shift and opposite phase shift in each MRM. For each, we calculate the RAMZI electrical field strength using the pre-determined arm phase shift. We then pick the combinations that yield RAMZI outputs with evenly-spaced electrical field strengths. This whole process up to this point is done at a very low laser power (-20 dBm input to each MRM) to make the effect of self-heating negligible. 

The final step is to raise the laser power to the desired value for circuit operation. This may make self-heating no longer negligible, so we sweep the MRM thermal tuning power again to find the value that yields the optimal biasing points (temperatures) found earlier. We rapidly oscillate the the driving voltage at MHz speeds during the sweep to mimic realistic MRM tuning and operation. Although the driving voltage will oscillate at 10s of GHz speeds in realistic application, MHz speeds are sufficient because they are faster than the MRM thermal time constant ($\sim$ ms).

We repeat the above process for a range of MRM input coupling gaps. For each, we measure the \omae{} in addition to the phase error, which is the maximum phase variations between any two of the four states.  We pick a final coupling gap to maximize the RAMZI's \omae{} while maintaining a sufficiently low quality factor. Final Q values of $\sim$3500 will support high-speed baud-rates of 50+ Gbd.

Our simulations show that after the arm phase shifter is properly tuned, the individual RAMZI can achieve an input-normalized \omae{} of 0.64 from four electric field transmission levels evenly spaced between 0.08 and 0.72 with a negligible phase error of $\text{$<$3}^\circ$. The I/Q eye-diagrams of a RAMZI-based offset-QAM-16 transmitter are shown for amplitude (a) and phase (b) at an example baud-rate of 50 GBd for 200 Gb/s transmission in Fig.~\ref{fig:eye diagram}.

The multiple steps required for simulation methodology will not impede the efficiency and practicality of the proposed transmitter in actual datacenter applications. First, the driving voltages for multi-level operation ($V_{MRM,T}, V_{MRM,B}$) will be the same across MRMs in different transmitters and not change over time, so they only need to be calculated once. Second, while device temperature and thus the required heating power can fluctuate in actual applications due to external sources (i.e. thermal coupling with co-packaged GPU), power-efficient feedback-based thermal alignment and tuning of MRMs has been well studied and demonstrated in previous works \cite{intel_thermal_tuning}. 

\section{Link Level modeling}
\label{sec:link modeling}

\begin{figure}[tb]
    \centering
    \includegraphics[width=0.9\linewidth]{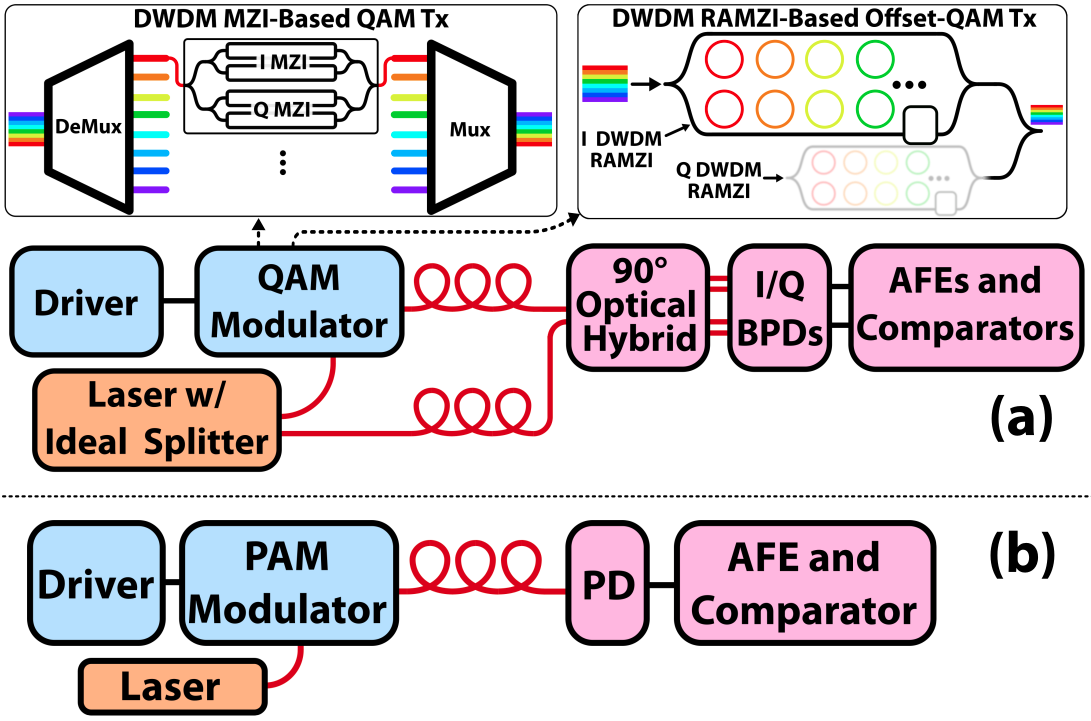}
    \caption{Coherent laser-forwarded (a) and IM-DD (b) link architectures used for modeling link-level BER. The MZI-based transmitters require an additional wavelength mux and demux for WDM operation (mux/demux not shown in part b).}
    \label{fig:link diagrams}
\end{figure}

In order to demonstrate the effectiveness of the ROQ-T and compare to state-of-the-art solutions, we simulate the link-level bit error rate (BER) as a function of total laser power for coherent and IM-DD formats with MZI- and MRM-based transmitters at target datarates of 200 and 400 Gb/s (per fiber per wavelength). For each we compare the performance of our ROQ-16 link to MZI-based PAM-4, PAM-8, QAM-4, and QAM-16 links, in addition to MRM-based PAM-4. System diagrams for QAM and PAM links are shown in Fig.~\ref{fig:link diagrams}.

For each combination of datarate and modulation format we calculate the appropriate baud rate, then calculate the peak-to-peak signal current at the receiver, and finally calculate the BER. The BER calculation takes into account peak-to-peak signal current, constellation geometry, receiver noise, and comparator threshold. The peak-to-peak signal current calculation takes into account coupler losses, link architecture, electro-optical modulator OMA/\omae{} and insertion loss, photodiode responsivity, optical mux/demux loss (only for MZI-based systems), and optical splitting/combination losses. We model the MRM with a 67 GHz bandwidth~\cite{MRM_bw} and the MZM with a 100 GHz bandwidth, even though the best MZIs reported in literature have bandwidths between 60-80 GHz~\cite{MZI_bw, silicon_2, InP_1, InP_2}. We assume all other circuit components to have infinite bandwidth. Full parameters used in modeling are shown in Tab.~\ref{tab:my-table}. We assume the MRM in the MRM-based PAM link to have an OMA of 0.4 and the RAMZI in the ROQ-16 link to have an \omae{} of 0.64, both based on simulation in the GF SPCLO process. We assume all MZIs to be fully driven (i.e. $\pi$ phase shift in each arm for QAM, and $\sfrac{\pi}{2}$ in each arm for PAM).

\begin{figure}[h]
    \centering
    \includegraphics[width=0.9\linewidth]{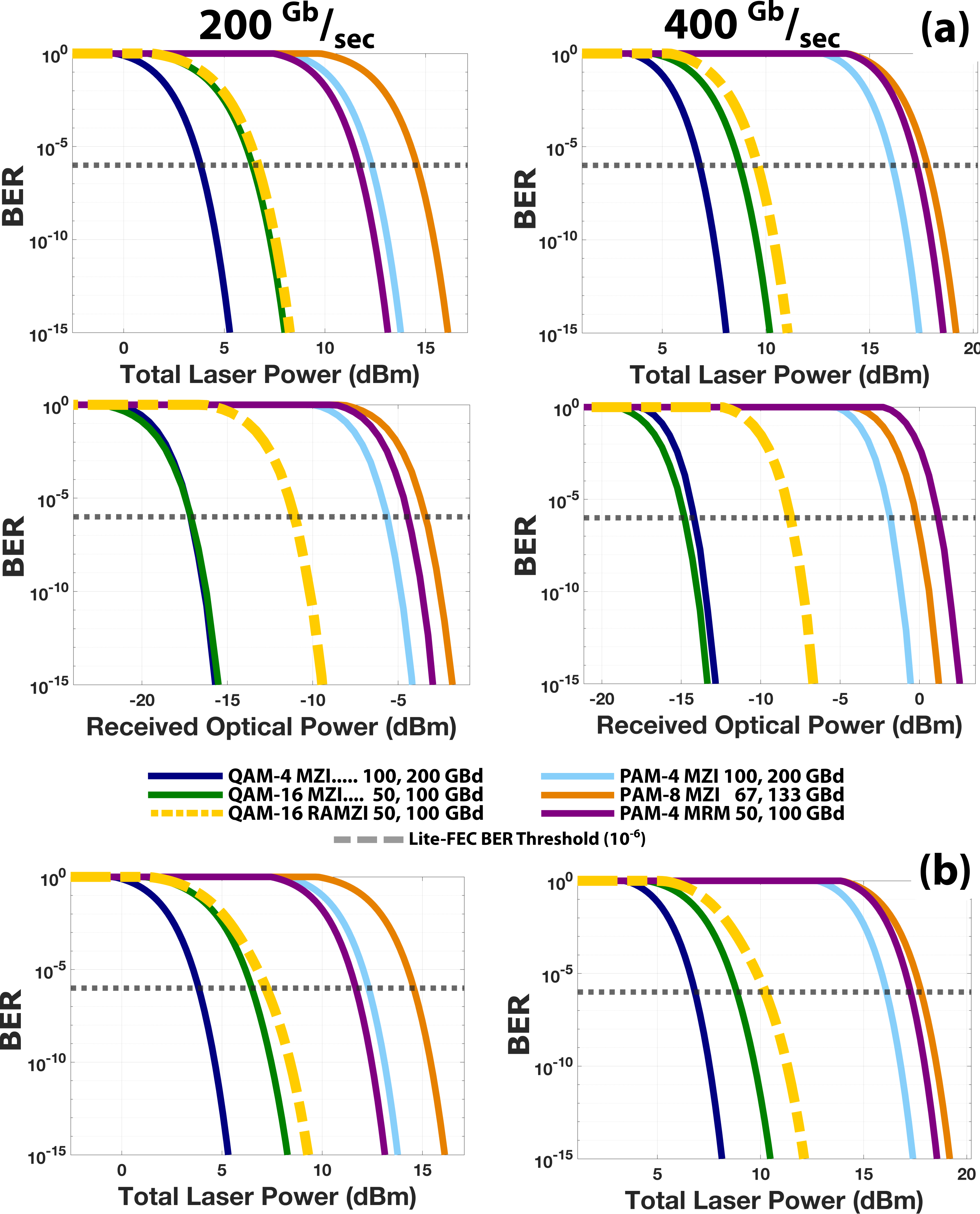}
    \caption{Simulated BER as a function of laser optical power for a single channel in a single direction with a) no phase noise, b) laser linewidth of 1 MHz and a path length mismatch of 1 cm.}
    \label{fig:ber laser power}
\end{figure}

To reflect the strict latency requirements of connected GPUs, we focus on optical power required to achieve a pre-FEC BER of $10^{-6}$. This has been adopted for the next generation of PCIe communications and supports a low latency ($\sim$ns)~\cite{pcie_6}. The modeling results show that the ROQ-16 link can achieve the target pre-FEC BER at a total optical power of 6.71 and 9.65 dBm for 200 and 400 Gb/s links (Fig.~\ref{fig:ber laser power}a). For both datarates this optical power is within 1 dB of that required by QAM-16 MZI-based links and 5.3 and 7.7 dB less laser power than that required by a link based on the current CPO solution of MRM-based PAM-4. Although the modeling shows that for both datarates, MZI-based QAM-4 will achieve the lite-FEC BER with lower optical power, the high baud rate it requires will lead to immense electrical power and make it impractical for high datarate links. 

To model the realistic operating environment of our proposed transmitter, we also simulate the effect of phase noise. Since in laser-forwarded CPO systems a single laser will be used to generate the LO and data signals used at the receiver, we anticipate that the path length mismatch between the two signals can be limited to 1 cm (due to fiber and on-chip waveguide mismatch). Based on commercially-available DFB O-Band lasers, we assume a laser linewidth of 1 MHz. These results are shown in Fig.~\ref{fig:ber laser power}b. Offset-QAM can be more susceptible to phase noise than non-offset-QAM because the offset-QAM constellation occupies a smaller range of the total I-Q phase space (Fig.~\ref{fig:offset qam constellation}). However, our simulation shows that even with phase noise taken into account, our ROQ-16 link can achieve the target pre-FEC BER at a total optical power within 2 dB of that required by QAM-16 MZI-based links. These results indicate the proposed link architecture using ROQ leads to 280~fJ/b laser energy-efficiency for both 200~Gb/s and 400~Gb/s (assuming 10\% laser wall-plug efficiency~\cite{Bowers2022-LaserComb, Intel2023-WDMLaser}).

\begin{table}[!htbp]
\centering
\resizebox{0.8\columnwidth}{!}{%
\begin{tabular}{@{}ccc@{}}
\toprule
\textbf{Parameter} & \textbf{Value} & \textbf{Citation} \\ \midrule

Grating Coupler Insertion Loss  & $1.8\,\mathrm{dB}$              & \cite{grating_coupler} \\ \midrule
Optical Mux (Demux) Loss        & $2.8\,\mathrm{dB}$              & \cite{mux_demux}        \\ \midrule
Extra Link Margin               & $5\,\mathrm{dB}$                &                         \\ \midrule
Photodiode Responsivity         & $1\,\sfrac{\mathrm{A}}{\mathrm{W}}$ & \cite{pd_resp}     \\ \midrule
MZI Insertion Loss              & $4\,\mathrm{dB}$                & \cite{mzi_il}           \\ \midrule
RAMZI E-Field Offset$^\dag$     & $0.42$                          & Sim.                    \\ \midrule
RAMZI \omae{}                   & $0.64$                          & Sim.                    \\ \midrule
MRM OMA                         & $0.4$                           & Sim.                    \\ \midrule
MZI \omae{}                     & $2$                             & \cite{google_imdd_coherent} \\ \midrule
MZI OMA                         & $1$                             & \cite{google_imdd_coherent} \\ \midrule
AFE Input-Ref Noise             & $15\,\sfrac{\mathrm{pA}}{\sqrt{\mathrm{Hz}}}$ & \cite{afe_noise} \\ \midrule
Input-Ref Comparator Threshold  & $\left(\sfrac{\text{Bd Rate}}{50\,\text{G}}\right)\times 5\,\mu\mathrm{A}$ & \cite{comp_threshold} \\ \midrule
MRM 3 dB Bandwidth              & $67\,\mathrm{GHz}$              & \cite{MRM_bw}           \\ \midrule
MZI 3 dB Bandwidth              & $100\,\mathrm{GHz}$             & \cite{MZI_bw}           \\

\bottomrule
\end{tabular}
}

\vspace{1.5mm}
\caption{Parameters used in link-level BER modeling.\newline
$^\dag$The RAMZI E-field offset refers to the mean output E-field amplitude, relative to RAMZI input E-field amplitude.}
\label{tab:my-table}
\end{table}

\section{Thermal Stability Discussions} \label{thermal analysis}

\begin{figure}[h]
    \centering
    \includegraphics[width=0.75\linewidth]{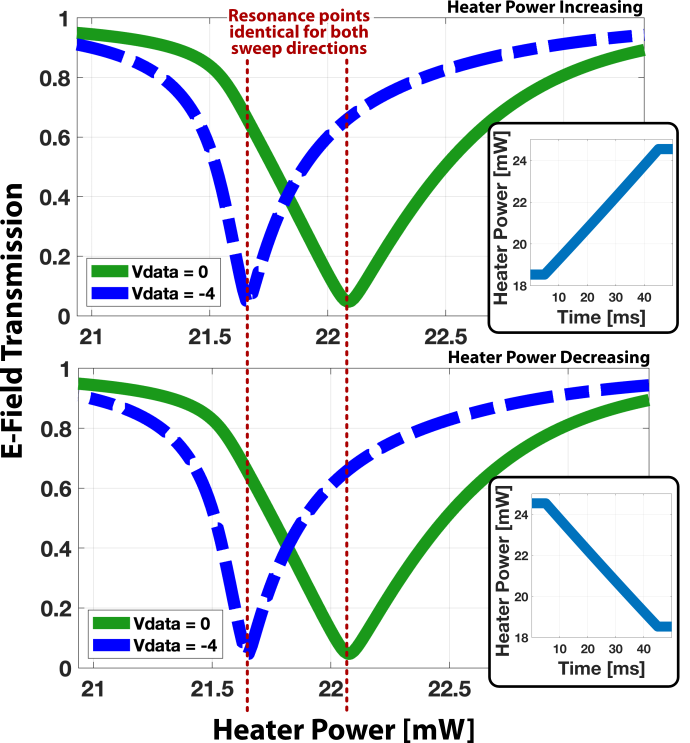}
    \caption{The E-Field normalized transmission of an MRM as a function of heater power for both increasing and decreasing power directions.}
    \label{fig:-5 dbm tuning}
\end{figure}

RAMZI operation requires that the two MRMs be thermally biased with resonances on both the left and right side of the laser wavelength (Fig.~\ref{fig:ramzi biasing}). This deviates from how MRMs are traditionally used as transmitters, in which their resonance is placed on only the right side of the laser wavelength to ensure thermal stability by making the feedback between optical power and device temperature negative~\cite{OptExp2019-RingPower}. In order to ensure the thermal-stability of the RAMZI circuit for 200 and 400 Gb/s links, we simulate the thermal performance of the MRM at the requisite optical powers in our Spectre-based simulation Virtuoso environment. 

To model a realistic MRM tuning application, we input various levels of optical power at a 1310 nm wavelength into a single MRM while rapidly applying 0V and -4V at 2 MHz. We sweep the thermal tuning voltage in both increasing and decreasing directions between 21 and 23.5 mW over 50 ms to modulate the resonance point of the MRM from below the laser wavelength to above it. These modulation speeds are chosen to model realistic MRM tuning by being higher (data modulation) or lower (thermal modulation) than the thermal time constant of the MRM ($\sim$ 1 ms). In the resulting MRM E-field transmission we separate the points coming from the applied voltages of 0V and -4V. We then examine the outputs for thermal stability. The first sign of thermal stability is no bistability - that the MRM output must have a one-to-one relationship with heater power with no dependence on the direction from which that power is approached (i.e. heating vs cooling). The second sign of thermal stability is no metastability - during the thermal sweep the MRM output must change smoothly and continuously with no abrupt jumps. 

Our simulation shows that the MRM features no thermal instability at input optical powers of less than or equal to -5 dBm (Fig.~\ref{fig:-5 dbm tuning}). The system-level optical powers of 6.7 and 9.7 dBm  required for low-BER RAMZI-based offset-QAM-16 operation (Sec.~\ref{sec:link modeling}, Fig.~\ref{fig:ber laser power}) equate to higher MRM input powers of -4.1 and -2.1 dBm. However, we believe that as power handling and dissipation of MRMs improve in coming years platforms (i.e. low loss waveguides in silicon), MRMs will be able to support these higher optical powers required for next-generation datarates of 200 and 400 Gb/s.

\section{Experimental Results} \label{chip results}

\begin{figure*}[h]
    \centering
    \includegraphics[width=0.95\textwidth]{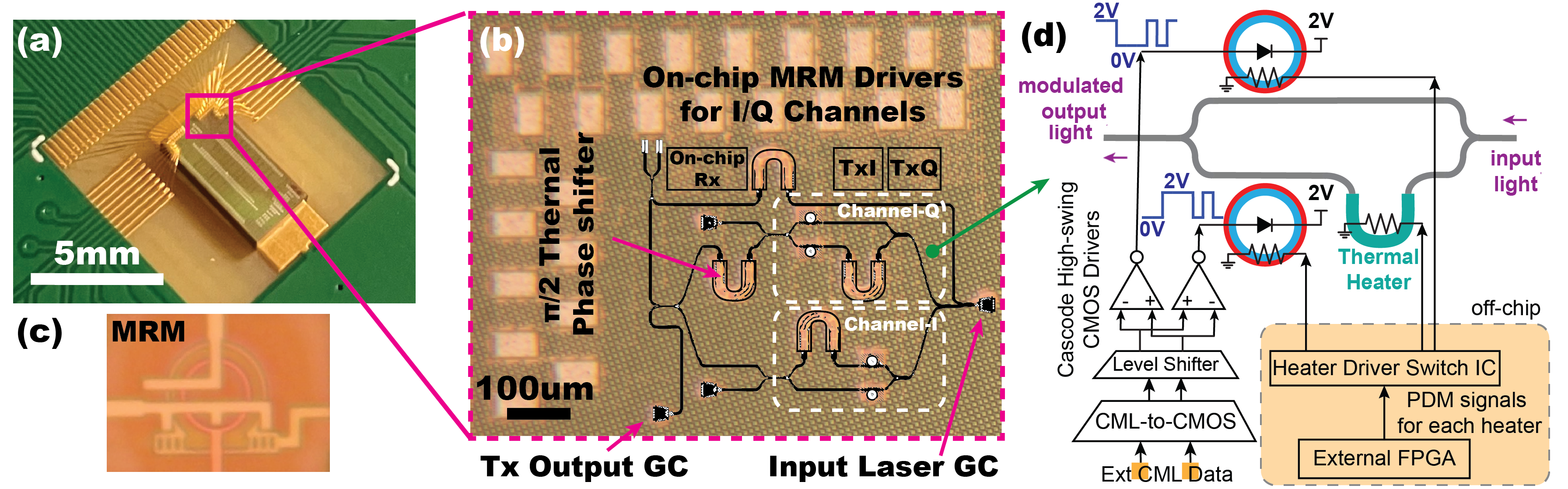}
    \caption{(a) Photo of wire-bonded chip fabricated in GF45SPCLO, (b) optical transmitter area overlapped with photonic circuits, (c) micrograph of an MRM, and (d) block-diagram of on-chip driver and control CMOS circuitry.}
    \label{fig:chip_photos}
\end{figure*}

\begin{figure*}[h]
    \centering
    \includegraphics[width=0.95\textwidth]{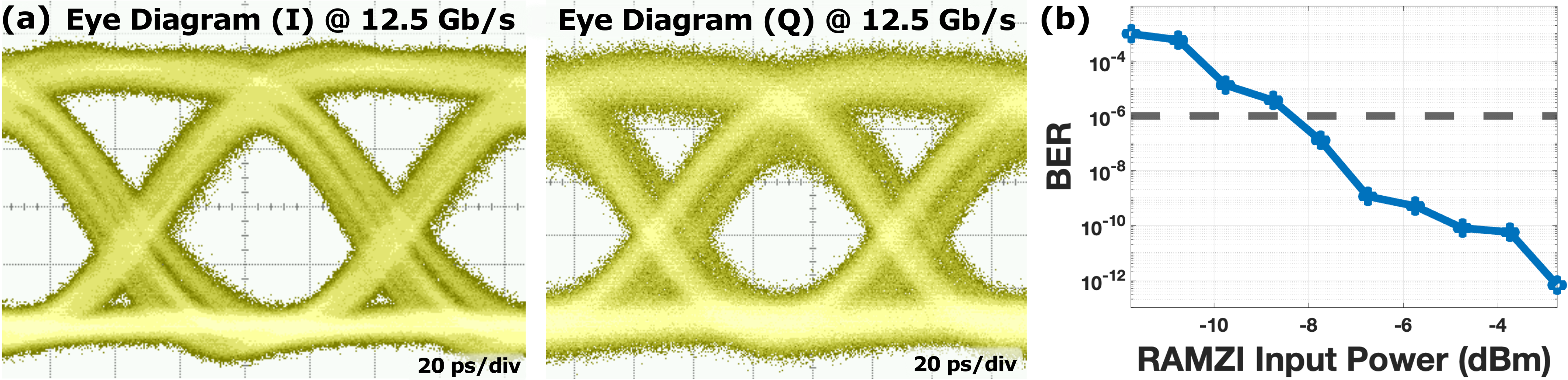}
    \caption{(a) Measured I/Q eye-diagrams, and (b) characterized BER curves vs. RAMZI input optical power.}
    \label{fig:meas_results}
\end{figure*}

A proof of concept of the proposed offset-QAM optical transmitter is fabricated and validated using the GlobalFoundries 45 nm monolithic silicon photonics process (GF45SPCLO)~\cite{ GF45SPCLO-OFC2020}. For an electronic transmitter, the chip has custom-designed analog front ends (AFEs) consisting of CML-to-CMOS conversion and high-swing drivers using a VDD supply of 1.1V to support a 2V peak-to-peak drive voltage for MRMs. All MRMs are biased in full depletion mode. Thermal phase shifters and MRM heaters are controlled via external pulse density modulation (PDM) signals at a 48 MHz clock frequency from an FPGA. To demonstrate coherent amplitude modulation, the chip is designed to perform QAM-4 and not QAM-16. 

The chip is wire-bond packaged directly on a PCB as shown in Fig.~\ref{fig:chip_photos}. The CML external data at 12.5 Gb/s was supported by a pattern generator and bit error rate test (BERT) equipment (Agilent J-BERT N4903A). Although all of the CMOS circuitry and MRMs can support up to 50 GBd, reported data rates are currently limited by wire-bond packaging and equipment capabilities. In addition to on-chip Tx circuitry, we also implemented an on-chip coherent receiver (Rx) front-end using balanced photodetectors (BPD), a trans-impedance amplifier (TIA), and a CML buffer for quick debugging and calibration. 

All optical circuits are designed to operate in the O-band (1310 nm), and MRMs have a radius of 7.5~$\mu$m with $\sim$45 pm/V modulation efficiency and a Q-factor of 3500. Light is coupled into and out of the chip using grating coupler (GC) devices with $\sim$4 dB loss in our current setup.


In order to tune all MRMs and phase shifters, we first swept the laser wavelength in the O-band and adjusted the MRMs thermally to separate their resonances by a fixed offset of $\sim$100 pm. We then applied the 12.5 GHz data and 2V MRM drivers and swept the laser to select the wavelength that yielded identical average power output in each MRM. We sensed MRM resonance points in this tuning process by connecting a high-precision voltage source to each MRM cathode to detect current proportional to optical power inside the MRM cavity. The final step was to sweep the power applied to the thermal phase shifter in each RAMZI and pick the value that maximized the observed OMA.

The I/Q eye diagrams at 12.5 Gb/s are shown in Fig.~\ref{fig:meas_results}a. We characterize the BER as a function of total laser power for a single channel and show that our RAMZI-based offset-QAM transmitter can achieve the lite-FEC threshold of $10^{-6}$ with a RAMZI optical power input of -8 dBm (Fig~\ref{fig:meas_results}b). These are all raw characterizations of the transmitter without using any equalization. The energy-efficiency is estimated to be 0.7 pJ/bit (including pre-drivers) for the overall 25 Gb/s offset-QAM-4 transmitter.

\section{Comparison to State-of-the-Art Optical QAM-16 Transmitters} \label{comparison SoA}

\begin{table*}[t]
\centering
\caption{Comparison of state-of-the-art high-speed optical transmitters.}
\label{tab:comparison_to_soa}

\resizebox{0.95\textwidth}{!}{%
{\tiny
\begin{tabular}{cccccccc}
\toprule

\textbf{\begin{tabular}[c]{@{}c@{}}Platform,\\ Work\end{tabular}} &
\textbf{Device} &
\textbf{\begin{tabular}[c]{@{}c@{}}Modulation\\ Format\end{tabular}} &
\textbf{\begin{tabular}[c]{@{}c@{}}Data Rate\\ \textnormal{($\sfrac{\textnormal{Gb}}{\textnormal{s}}$)}\end{tabular}} &
\textbf{\begin{tabular}[c]{@{}c@{}}Baud Rate\\ \textnormal{(GBd)}\end{tabular}} &
\textbf{\begin{tabular}[c]{@{}c@{}}Area\\      \textnormal{($\text{mm}^2$)}\end{tabular}} &
\textbf{\begin{tabular}[c]{@{}c@{}}Bandwidth\\ \textnormal{(GHz)}\end{tabular}} &
\textbf{\begin{tabular}[c]{@{}c@{}} $V_{\pi}L$ Efficiency  \\ \textnormal{(V $\cdot$ cm)}\end{tabular}} \\

\midrule
SOI~\cite{silicon_1} & MZI & QAM-16 & 400   & 120   & 3.5$^\dag$     & 37.7 & NA     \\
SOI~\cite{silicon_2} & MZI & QAM-64 & 556.8 & 116   & 4$^\dag$       & 63   & 3.6     \\
TFLN~\cite{TFLN_1}   & MZI & QAM-32 & 533   & 128   & 23$^\dag$      & 60   & 2.9   \\
InP~\cite{InP_1}     & MZI & QAM-16 & 400   & 128   & NA             & 64   & NA     \\
InP~\cite{InP_2}     & MZI & QAM-16 & 576   & 144   & 12.5$^\dag$    & 90   & 2.5     \\
SOI~\cite{silicon_mrm_china} & MRM & PAM-8  & 240   & 80    & 3.4e-3       & 110  & 0.825  \\
SOI~\cite{silicon_mrm_intel} & MRM & PAM-4  & 240   & 120   & 6.4e-5       & 54   & 0.53   \\
This Work$^\ddag$     & MRM & QAM-16 & 400   & 100   & 2e-2           & 110  & 0.825  \\
This Work$^\P$        & MRM & QAM-4  & 25    & 12.5  & 0.2            & NA   & 0.5    \\
\bottomrule
\end{tabular}
} 
} 

\vspace{1ex}
\begin{minipage}{0.95\textwidth}
\raggedright
$^\dag$ Area is estimated from reported device length by assuming an MZI width of $0.5\;\text{mm}$.\\
$^\ddag$ Theoretical analysis: bandwidth and efficiency obtained from reported literature on MRM-based transmitters. Area estimate based on a 4-$\lambda$ WDM RAMZI-based QAM Tx developed in the GF 45 nm SPCLO process in which one thermal phase shifter is shared for all MRMs in each I/Q modulator.


$^\P$ Experimental results.
\end{minipage}
\end{table*}

To demonstrate the energy and area efficiency of the ROQ-T relative to conventional approaches we compare out work to state-of-the-art MZI-based QAM transmitters made from Silicon, Indium Phosphide, and Thin-Film Lithium Niobate (TFLN) in Tab.~\ref{tab:comparison_to_soa}. Since our fabricated chip was a proof of concept not optimized for CPO environments, we estimate the performance characteristics of our proposed transmitter based on reported results from literature on MRMs and MRM drivers for PAM-4 operation.

State-of-the-art 4-level (PAM-4) MRM drivers have been demonstrated with efficiency on the order of 50 fJ/bit~\cite{sajjad_mrm_driver}. Since the ROQ-T will require two MRMs per dimension instead of one as is used for amplitude modulation, the efficiency will be roughly 100 fJ/bit. Meanwhile, MZM drivers for higher-order QAM operation have been demonstrated with efficiencies of 2-3 pJ/bit~(\cite{MZM_driver_palermo, MZM_driver_japan}). 

Our work thus has a comparable power efficiency and datarate to MZM-based QAM-16 transmitters while providing more than 10-100$\times$ lower area. Compared to PAM-4 MRM-based transmitters, our work has similar area and electrical power efficiency but double the datarate per wavelength without increasing the baud rate or number of WDM laser lines. It also requires less laser power than PAM-4 MRM-based transmitters due to the higher sensitivity of coherent receivers (Fig.~\ref{fig:ber laser power}).

\section{Future Work and Conclusion} 
\label{conclusion}

Co-packaged optics (CPO) is becoming imminent to meet the immense and growing demands of GPUs and network switches for rapid, dense, and energy-efficient optical intra-datacenter interconnect. However, shoreline density and thus overall datarates will be limited by the scaling of polarization and wavelength-division multiplexing as well as an ability of electronics to run at higher frequencies. Achieving target datarates for future datacenters of 400+ Gb/s per fiber per wavelength with high energy efficiencies can be achieved by using microring modulators for more advanced and coherent modulations. 

In this work, we propose and demonstrate the usage of an O-band frequency-detuned ring-assisted Mach-Zehnder interferometer (RAMZI) for phase-constant amplitude transmission along with a RAMZI-based offset-QAM transmitter (ROQ-T). We show its performance in a circuit-level simulation with devices from a leading silicon photonics process, achieving an input-normalized effective optical modulation amplitude of 0.64 per dimension while holding phase error below 3$\degree$. We conduct a full link-level modeling to show that it can achieve the target datarates with an optical power nearly identical to that required by conventional MZI-based modulators despite consuming $\sim$10-100$\times$ less area. Additionally, we simulate its thermal properties to show its feasibility for operation at the required optical power. Finally, we demonstrate a proof of concept of our transmitter in the Global Foundries 45 nm silicon photonics process performing 25 Gb/s offset-QAM-4.

In the future we plan to demonstrate the presented approach across multiple channels at higher baud rates using on-chip clocking and pattern generation. Additionally, we will implement a closed-loop thermal sensing and control system to both align and lock MRM resonances to proper offsets on both sides of the laser wavelength, as is required for RAMZI operation. We have recently shown that offset-QAM modulation can be used to realize DSP-free, low-latency and energy-efficient C2PO link architectures~\cite{marizyeh_recovery}, and will demonstrate this physically as well. With these future developments we will further demonstrate that RAMZIs and RAMZI-based offset-QAM transmitters will provide a compact and energy-efficient solution for CPO systems to continue scaling to meet the growing demands of GPUs, network switches, and datacenter interconnect at sustainable energy efficiencies.


\bibliography{sample}
\end{document}